\documentclass[conference,compsoc]{IEEEtran}
\ifCLASSOPTIONcompsoc
  \usepackage[nocompress]{cite}
\else
  \usepackage{cite}
\fi
\usepackage{hyperref}

\ifCLASSINFOpdf
\else
\fi

\usepackage{datetime2}
\usepackage{xcolor,bm} 
\usepackage{enumerate,enumitem} 

\usepackage{booktabs,multirow} 
\usepackage{amssymb}
\usepackage{amsmath,amsfonts,amsthm, mathtools} 
\usepackage{subcaption} 

\usepackage{color}
\usepackage{tikz} 

\newcommand{\beqa} {\begin{eqnarray}}
\newcommand{\eeqa} {\end{eqnarray}}

\newcommand{\bra}{\langle}
\newcommand{\ket}{\rangle}

\begin{document}
%
\title{QDCNN: Quantum Dilated Convolutional Neural Network}

\author{\IEEEauthorblockN{Yixiong Chen}
\IEEEauthorblockA{IBM Corporation\\
Email: cnyixg@cn.ibm.com}
}


%


\maketitle

\begin{abstract}
In recent years, with rapid progress in the development of quantum technologies, quantum machine learning has attracted a lot of interest. In particular, a family of hybrid quantum-classical neural networks, consisting of classical and quantum elements, has been massively explored for the purpose of improving the performance of classical neural networks. In this paper, we propose a novel hybrid quantum-classical algorithm called quantum dilated convolutional neural networks (QDCNNs). Our method extends the concept of dilated convolution, which has been widely applied in modern deep learning algorithms, to the  context of hybrid neural networks. The proposed QDCNNs are able to capture larger context during the quantum convolution process while reducing the computational cost. We perform empirical experiments on MNIST and Fashion-MNIST datasets for the task of image recognition and demonstrate that QDCNN models generally enjoy better performances in terms of both accuracy and computation efficiency compared to existing quantum convolutional neural networks (QCNNs).
\end{abstract}

\maketitle

\section{Introduction}\label{introduction}
Convolutional neural networks (CNNs), proposed by Yann LeCun et al \cite{lecun1998gradient} in 1989, are one of the most powerful algorithms in the context of deep learning. The main advantage of CNNs is that they use multiple
feature extraction stages to automatically and accurately learn important features from the data without any human supervision. Due to this advantage, CNNs have been tremendously successful in a broad array of high-level computer vision problems, including image recognition \cite{krizhevsky2012imagenet, simonyan2014very, szegedy2015going, he2016deep} , object detection \cite{girshick2014rich, girshick2015fast, redmon2016you}, and image segmentation \cite{he2017mask, ronneberger2015u, chen2018encoder}. In recent years, with further development in deep learning , CNNs have also been demonstrated to show promising performances in other machine learning areas such as time series forecasting \cite{borovykh2017conditional, chen2020probabilistic}, speech recognition \cite{45774} and recommendation system \cite{yuan2019simple}. 

Parallelly, with recent achievements in quantum technologies (e.g. noisy intermediate-scale quantum (NISQ)
processors are currently availlabe), the domain of quantum machine learning has attracted growing concerns and triggered an enormous amount of work. Quantum machine learning is a research area with the purpose of utilizing quantum mechanical effects such as superposition and entanglement to improve the performance of machine learning algorithms. Even though quantum machine learning is a new discipline, it has witnessed a number of successful quantum extensions to classical machine learning problems, including support vector machine \cite{varatharajan2018big}, clustering \cite{kerenidis2018q,otterbach2017unsupervised},  and principal component analysis \cite{article}.

Among quantum machine learning algoritms, quantum convolutional neural networks (QCNNs), also known as hybrid quantum-classical convolutional
neural networks, are a family of variational quantum algorithms and they have recently become a very active research field. The central idea of QCNNs is to construct a quantum convolutional layer within neural networks based on parameterized quantum circuits to estimate complex kernel functions in high dimensional Hilbert space. Inspired by CNNs, Liu et al. \cite{liu2019hybrid} proposed the first QCNN model and implement it for image recognition. Afterwards, the QCNN model was investigated further in various work \cite{cong2019quantum, henderson2020quanvolutional, oh2020tutorial, chen2020quantum, houssein2021hybrid, alam2021iccad}. Recently,  it has been demonstrated in \cite{yang2021decentralizing} that QCNN models can also achieve promising results in speech recognition. 

Despite these successes, QCNNs suffer from computational bottlenecks which make it time consuming to train QCNNs. Firstly, quantum operations applied on n-qubit quantum circuits require unitary matrices of size $2^n\times2^n$ which will scale exponentially as the size of quantum circuit. Moreover, the calculation of gradients, due to the parameter-shift rule \cite{mitarai2018quantum, schuld2019evaluating}, result in more quantum circuit executions, when QCNNs are trained on a real quantum device. As an example, a quantum filter with $p$ trainable parameters will add $2p$ more quantum circuit executions for each training sample to compute the required gradients. Even though this problem can be mitigated when QCNNs are implemented on quantum simulators that support more efficient gradient computation methods such as back-propagation \cite{Linnainmaa:1976,Rumelhart:1986we} and adjoint differentiation method \cite{jones2020efficient}, it is inevitable for QCNNs to face another challenge. In CNNs, a convolutional layer, due to local connectivity,  performs a large amount of element-wise matrix multiplication operations. For example, an output $100\times 120 $ feature map of a convolutional layer is obtained from $100\times 120 = 12,000$ multiplication operations. The computational cost will increase significantly with the feature map size. Fortunately, this computational issue in CNNs can be handled by using vectorization techniques \cite{Chellapilla_highperformance, vectorization}. QCNNs, as the counterpart of CNNs, have the same problem. However, unlike CNNs, most of current quantum devices, including quantum hardware and quantum simulator, do not support vectorization. Despite the availability of more mature quantum devices in the NISQ era, executing a large number of quantum circuits would be impractical in general.

A few works have been done to investigated how to reduce the runtime complexity of QCNNs. In the first family of works, a small number of qubits required for the quantum circuit is achieved by  using classical data pre-processing techniques to reduce the dimension of the input features fed into the quantum (convolutional) layer. For instance, Pramanik et al. \cite{pramanik2021quantum} employ  principal component analysis (PCA) to reduce the VGG-16 features for the quantum variational classifier (VQC), while Hur et al. \cite{hur2021quantum} adopt autoencoding (AutoEnc) for the dimensionality reduction. Nevertheless, the performance of the model trained in this way is likely to be compromised by the limited expressive power of the reduced features, as shown in \cite{pramanik2021quantum}. The second family of works focus on how to efficiently encode classical data into quantum states. Schuld and Killoran  \cite{schuld2019quantum} propose and implement the \emph{amplitude encoding} for variational quantum circuits, which is explored further in \cite{mattern2021variational} for Flexible Representation of Quantum Images (FRQI).  This type of encoding method is efficient in terms of required qubits for data encoding but it relies on too deep quantum circuits which are unpractical on NISQ devices.  In a different direction,  some recent researchers \cite{schuld2018supervised, larose2020robust, alam2021iccad} propose angel encoding (also referred to as qubit encoding) and its variants (e.g. dense angel encoding) which use a constant quantum circuit depth for state preparation. This encoding scheme requires one qubit to encode one or a limited number of components of the input feature vector and thus is not efficient for high-dimensional input features from a resource prospective. 
To trades off these two encoding methods mentioned above,  Hur et al. \cite{hur2021quantum}  further develop a hybrid encoding approach which requires fewer number of qubits
than the angel encoding and use shallower quantum circuit depth than the amplitude encoding. Moreover, Henderson et al. \cite{henderson2020quanvolutional} employ a threshold based encoding technique to reduce the input-state space and made it possible to obtain the output feature map through a look-up table during the quantum convolution process without needing to execute the same quantum circuit repeatedly on image segments. This method is easy to implement, but it is infeasible on real quantum devices,  as mentioned in \cite{henderson2020quanvolutional}.  

Having reviewing all these challenges and developments, in this work, we propose a novel hybrid quantum-classical architecture which we will call \emph{quantum dilated convolutional neural network} (QDCNN). Our approach, motivated by the dilated convolution in deep learning, is an extension of the architectures presented in \cite{liu2019hybrid} and \cite{henderson2020quanvolutional}, and helps reduce the computational cost of QCNNs in a different way compared to the aforementioned approaches.
Dilated convolution, also known as atrous convolution, was originally developed for efficiently computating the undecimated discrete wavelet transform \cite{Holschneider1989}. In recent years, dilated convolution has attracted more and more attention, and is widely used in semantic segmentation \cite{Yu2016MultiScaleCA, Chen2015SemanticIS, chen2017deeplab, chen2017rethinking, chen2018encoder, hamaguchi2018effective}. Following these successes, dilated convolution has also been adopted for a broader set of tasks, such as object localization \cite{kudo2017dilated},  time series forecasting \cite{borovykh2017conditional,chen2020probabilistic} and sound classification \cite{chen2019environmental}.
The advantage of dilated convolution is that it  allows for  effectively expanding the field of view of filters to capture larger context without increasing the number of parameters or the computational complexity. By virtue of dilated convolution, the proposed QDCNNs can generally improve the computational efficiency of existing QCNNs while achieving the better task performance.

In summary, the contributions of our work are \vspace{0.5em}

\begin{itemize}
\setlength\itemsep{1em}
	\item We propose a novel architecture of quantum convolutional neural network based on quantum dilated convolution operation. To the best of our knowledge, our work is the first attempt to combine  the concept of dilated convolution with variational quantum circuits.
	\item We conduct experiments using MNIST and Fashin-MNIST datasets and demonstrate the superior performance of QDCNN models over QCNN models.
\end{itemize}

\section{Method}\label{method}
\subsection{Preliminaries}

\subsubsection{Convolution Operation}   
 The convolutional layer, which performs an operation called a “convolution“, plays a central role in CNNs. In the context of convolutional networks, a convolution is a linear operation that involves the multiplication of a set of weights with the input. For a convolution operation, a \emph{kernel} or \emph{filter} is defined as a feature extractor which is a two-dimensional (2-D) array of learnable weights. A filter is applied to a filter-size patch of the input image called \emph{receptive field} and a dot product is performed between the pixels within the receptive field and the weight values in the filter. Afterwards, the filter shifts to the next patch according to a step size called \emph{stride}, and repeats the above process until it has swept across the entire image. The final output from the series of dot products  between the filter weights and the values underneath the filter, is called a \emph{feature map}. Let us denote the output feature map by $y$ and the input image by $x$. In the 2-D convolution process, the feature map $y$ is obtained by applying a filter $k$ to the input image $x$:
\beqa \label{conv}
y[i,j] = \sum_q \sum_l k[q,l] \cdot x[i+q, j+l]
\eeqa
where $i$ and $j$ are location indices of $y$. The output feature map, due to the convolution operation, usually has smaller spatial resolution than the input image. This reduction in dimensions can be avoided by employing \emph{zero padding} technique, namely adding a border of pixels with value zero around the edges of the input image before the application of a filter. A hyperparameter called \emph{padding} can be defined to determine how many zero values to add to the border of the image. Generally, the spatial resolution $o_w$ and $o_h$ of the resulting feature map, extracted from an $i_w \times i_h$ input image by a $m \times n$ kernel, can be calculated as
\beqa 
o_w &=& \left( \frac{i_w-m+2p}{s}\right) +1 \label{output_dim1} \\
o_h &=& \left( \frac{i_h-n+2p}{s}\right) +1 \label{output_dim2}
\eeqa
where $p$ and $s$ represent padding and stride respectively.

\subsubsection{Dilated Convolution}  Dilated convolution is a type of convolution that expands the kernel by inserting holes (i.e. points with weight of zero) between the consecutive kernel elements. In simple terms, dilated convolution is just a convolution applied to the input with defined gaps. Compared to standard convolution, dilated convolution introduces an extra hyperparameter called \emph{dilation rate} that determines the stride with which the input pixels are sampled. According to the definition of dilated convolution,  $r-1$ zero values are inserted between two consecutive filter values, if the dilation rate is denoted by $r$. In this spirit,  Eq. \eqref{conv} needs to be reformulated as 
\beqa \label{dilated_conv}
y[i,j] = \sum_q \sum_l k[q,l] \cdot x[i+q \cdot r, j+l\cdot r]
\eeqa
in the context of dilated convolution.
It can be seen from Eq. \eqref{dilated_conv} that dilated convolution is able to capture a larger receptive field without introducing more learnable parameters compared to standard convolution with the same kernel size.  Moreover, for dilated convolution, we also need to rewrite Eq. \eqref{output_dim1} and Eq. \eqref{output_dim2} as 
\beqa 
o_w &=& \left( \frac{i_w-m - (m-1)(r-1)+2p}{s}\right) +1 \label{output_dim1_dilated} \\
o_h &=& \left( \frac{i_h-n- (n-1)(r-1)+2p}{s}\right) +1 \label{output_dim2_dilated}
\eeqa
which indicate that dilated convolution generally results in a feature map with smaller size compared to standard convolution for the same set of hyperparameters. It is worth noting that the standard convolution can be regarded as a special case of dilated convolution with dilation rate $r=1$.

\subsubsection{Quantum Convolution} \, In contrast to classical convolution, quantum convolution is a new type of convolution based on quantum circuits and it generally consists of three modules:
\begin{itemize}
\item \emph{ENCODING MODULE}. \,In this module, classical data is encoded into a quantum state which will be further processed in the quantum convolutional circuit. There exist various encoding methods such as angle encoding, amplitude encoding and basis encoding. A summary of them can be found in the literature \cite{dataencoding}. Among these methods, angle encoding  is the most commonly used encoding approach. In this encoding scheme, the classical input is treated as the rotation angle of a single-qubit rotation gate (e.g. $RY$ rotation gate). For example, a classical variable or feature $a$ can be encoded by $RY(a)$ which is applied on some initial state (e.g. vacuum state $|0\ket$). In this sense, we can say that the classical information $a$ is encoded into the initial state of a qubit. This type of angle encoding is called \emph{one variable/qubit encoding}. This  approach requires $n$ qubits to encode $n$ input variables. To reduce the required qubits, we can also encode multiple variables by sequential rotations applied on a single qubit. For example, input variables $a_1$, $a_2$ and $a_3$ can be encoded using  $RX(a_1)$, $RY(a_2)$ and $RZ(a_3)$ rotation gates applied successively on a single qubit. This angle encoding is called \emph{multiple variables/qubit encoding} or \emph{dense angel encoding}. In this paper, we focus on one variable/qubit encoding method. Let us denote by $E(x)$ the encoding operator where $x$ is the input vector. Then the encoded quantum state is obtained by
\beqa
|x\ket = E(x) |0\ket.
\eeqa
It is worth noting that $E(x)$ usually contain the Hadmard gate which transforms the initial state into a superpostion state.
 \\

\item \emph{ENTANGLEMENT MODULE}.\, In this module, a cluster of single- and multi-qubit gates are applied to the encoded quantum state obtained from the previous module. Mutli-qubit gates are usually $CNOT$ gate and parametric controlled rotation gate (e.g. $CRZ(\theta)$ where $\theta$ is a trainable parameter), and they are used to generate correlated quantum states, namely entangled states. Single-qubit gates are mainly parametric rotation gates. This combination of  single- and multi-qubit gates is referred to as parameterized layer in a QCNN and is designed to  extract task-specific features. This parameterized layer is usually repeated multiple times to extend the feature space. If we denote all unitary operations in the entanglement module by $U(\theta)$ for simplicity, the output quantum state will be
\begin{equation}
	|x,\theta\ket = U(\theta) |x\ket.
\end{equation}\vspace{0.001cm}
\item \emph{DECODING MODULE}.\, At this stage,  certain local observable $A^{\otimes m}$ (e.g. Pauli Z operator $\sigma_z^{\otimes m}$) is measured in the final quantum state $|x,\theta\ket$ from the entanglement module, where $m$ is equal or smaller than the total number of qubits $n$ in the quantum system.  The expectation value of the chosen
observable $A^{\otimes m}$ can be obtained by repeated measurements:
\beqa
\label{expv}
f(x,\theta) = \bra x,\theta| A^{\otimes m} |x,\theta\ket 
\eeqa
So the purpose of this layer is to extract a classical output vector $f(x,\theta)$ by using the mapping from the quantum state to a classical vector:
\beqa
\mathcal{M}: \quad  |x,\theta\ket \rightarrow f(x,\theta).
\eeqa
This classical vector $f(x,\theta)$ can be used as the input features for the subsequent layer in the QCNN.
\end{itemize}

\begin{figure*}[htp!]
\centering
\includegraphics[width=0.95\textwidth]{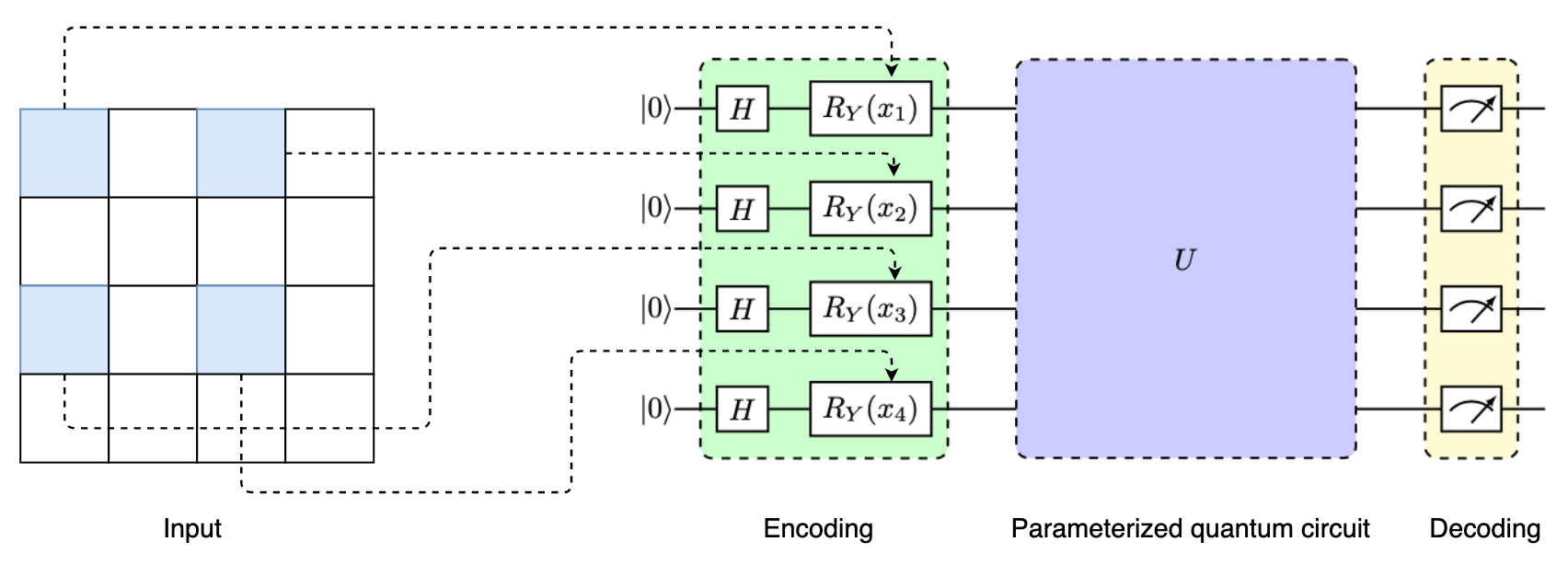}
\caption{An example of a quantum dilated convolutional layer (QDC) with kernel size $=2\times2$ and dilation rate $=2$. In contrast to standard quantum convolution, dilated convolution is applied to the input with defined gaps (one gap in this examle) to enlarge the receptive field. Encoding, entanglement, decoding modules for the QDC layer are highlighted with green, purple and yellow colors respectively.}
\label{qdc_layer}
\end{figure*}

\subsection{QDCNN}\label{qdcnn}

The proposed QDCNN is designed in the same fashion as QCNNs described in literatures \cite{liu2019hybrid, henderson2020quanvolutional}. Our model integrates quantum layers with classical layers and the quantum circuit ansatz can be placed anywhere in the model (e.g. at the beginning of the network, at intermediate layers in the network). 

The key difference between our method and existing QCNNs is that the dilated convolution is employed for the quantum convolutional layer. So the quantum layer in QDCNNs is called \emph{quantum dilated convolutional (QDC) layer}. An example of a QDC layer is illustrated in Fig. \ref{qdc_layer}. Due to the mechanism of dilated convolution , the quantum kernel in our model generally covers larger image patches (i.e. receptive fields ). For example,  a $2\times2$ quantum dilated convolution with dilation rate of 3 has a receptive field of $4\times4$ while the standard quantum convolution with the same kernel size has only a receptive field of $2\times2$ . It is noteworthy that even though the quantum dilated convolution is able to expand the receptive field the number of data points that are fed into the quantum convolution circuit is the same as the one for the standard quantum convolution. This means that the quantum dilated convolution does not requires more qubits than the standard quantum convolution with the same kernel size.

Our QDCNN model has mainly two advantages. Firstly, the QDC layer in our model, thanks to the enlarged receptive field, requires less number of times that the quantum kernel slides across the image (if there is no padding and the stride is the same), compared to the existing QCNN models. This can be understood by comparing Eqs. \eqref{output_dim1} and \eqref{output_dim2} with Eqs. \eqref{output_dim1_dilated} and \eqref{output_dim2_dilated} respectively.   Therefore, using the QDC layer helps reduce the number of quantum circuit executions during the quantum convolution process.  In the NISQ era, long training time is one of the biggest challenges facing the QCNN models. This difficulty mainly stems from the large number of quantum circuit executions from the quantum layers. In the quantum feature mapping process, due to the probabilistic characteristics, quantum measurement is usually performed multiple times (e.g. 1024) to get expectation values of some observables which can be considered as the extracted quantum feature maps. So how to reduce the number of quantum circuit executions plays a crucial role in mitigating the long-running-time problem of QCNNs. Our proposed quantum dilation is a powerful tool to explicitly control the amount of quantum circuit executions in the quantum layer.

The second advantage of our model is that it can improve the performance (e.g. classification accuracy) of existing QCNN models. Due to the expanded receptive field, the QDC layer in our model generally reduce the spatial resolution of the resulting feature maps. However, these feature maps are extracted from larger receptive fields of the image and hence contain long-range context which plays an essential role in many machine learning tasks such as image recognition and image segmentation.

\section{Experiments}\label{experiment}
In this section, we conduct two experiments to evaluate the performance of our proposed QDCNN model and compare it with the existing QCNN model. In \emph{Experiment A} and \emph{Experiment B}, we construct quantum convolutional models with non-trainable and trainable quantum filters, respectively.

\subsection{Experiment settings}
\subsubsection{Dataset} 
We choose the image benchmark MNIST and Fashion-MNIST datasets \cite{deng2012mnist, xiao2017fashion} for our experiments. The MNIST dataset contains 10 different classes of handwritten digits from `0' to `9' , while the Fashion-MNIST dataset is a collection of 10 different shapes of t-shirts, dresses, shoes, etc. Both of these datasets have 60,000 training samples, and 10,000 test samples of 28-by-28 gray scale  pixel images. Due to the expensive training and validation, we pick two subsets of the entire MNIST and Fashion-MNIST datasets, respectively, both of which consist of 1,000 balanced training samples and  200 balanced testing samples.

\subsubsection{Tested Models}
In this research, we consider two types of models:
\begin{itemize}
\item \emph{QDCNN Model.}  We employ the architecture of the most basic convolution-inspired hybrid quantum-classical neural network. Our QDCNN model consists of one QDC layer with one filter and one fully-connected layer with 10 neurons.
The kernel size and stride for the QDC layer is selected as $2\times2$ and 2 respectively without specification.  The quantum circuit ansatz of the QDC layer is designed as below. The 1 variable/qubit encoding scheme is adopted to encode the input image. Specifically, $2\times2$ pixels are encoded into a 4-qubit state using RY rotation gates. Note that, these $2\times2$ pixels are not adjacent to each other in the input image, due to the quantum dilated convolution. The resulting 4-qubit state is further transformed by a following random parameterized quantum circuit which might creates the entanglement. The decoding method  follows the same spirit of \cite{Pennylane-QCNN}, in which each expectation value is mapped to a different channel of a single output pixel. Consequently, even though there is only one filter, the quantum layer can transform the input 2-D image into four feature maps. This type of quantum layer might benefit the model performance as it allows for correlation among channels of the output feature maps.  In both cases of MNIST and Fashion-MNIST datasets, the QDC layer extracts from the $28\times28$ input  image a feature tensor of size $13\times13\times4$, which is then transformed to 10 output probabilities by the fully-connected layer with softmax activation. To evaluate how the dilation rate impacts the model performance, we consider two QDCNN models with dilation rate $r=2$ and $r=3$. We refer to these two models as \emph{QDCNN\_r2} and  \emph{QDCNN\_r3} respectively for the rest of the paper. 
\\
\item \emph{QCNN Model.} We choose the standard QCNN model as our benchmark model. The QCNN model follows the same structure of our QDCNN model with the only difference that it uses standard quantum kernel rather than dilated quantum kernel.
\end{itemize}
The random quantum circuit in each of these models consists of two 4-qubit random layers, each of which has four non-trainable or trainable parameters. For fair comparison, all of these random circuits share the same architecture generated by the same random seed.

\subsubsection{Training Setup} 
In Experiment A, after applying the non-trainable quantum filter to transform the original image data into feature maps, we use a mini-batch of 32 and Adam optimizer with a learning rate of 0.01 to train each model for 30 epochs. In Experiment B, due to the computational cost of training parametric quantum circuits involved in the trainable quantum filter, we reduce the batch size to four and train all models for 20 epochs with other hyperparameters remaining unchanged.

\subsubsection{Experimental Environment}
Experiments are conducted on the local computer with a 6-core CPU (2.2 GHz) by using PennyLane \cite{bergholm2018pennylane}, Qulacs \cite{suzuki2011qulacs} and PyTorch \cite{paszke2019pytorch}. PennyLane is an open source python-based framework that enables the automatic differentiation for hybrid quantum-classical computations. It is compatible with mainstream machine learning frameworks such as TensorFlow \cite{abadi2016tensorflow} and PyTorch, and it has a large plugin ecosystem which offers access to numerous quantum devices (i.e. simulators and hardware) from different vendors including IBM, Google, Microsoft, Rigetti and QunaSys.  In Experiment A, we perform the quantum processing of the original image data by using the \emph{Qulacs} simulator \cite{Qulacs} which is a high-performance C++ quantum simulator and made available through the community contributed PennyLane-Qulacs plugin \cite{Pennylane-Qulacs}. In Experiment B, considering the large amount of quantum circuit executions required in the scheme of parameter-shift rule, we train all hybrid models by using instead the built-in Pennylane simulator \emph{default.qubit} which supports back-propagation method for PyTorch interface. 
\subsection{Results}

As demonstrated in Table \ref{table1} and Table \ref{table2}, our proposed QDCNN models exhibit significant performance benefits over the QCNN model in terms of run-time efficiency.  More specifically, QDCNN models in Experiment A and B speed up the model training process by up to 15.24\% and 18.42\% respectively, compared with the QCNN model (QDCNN\_r2 and  QDCNN\_r3 have similar training time because their QDC layers output feature maps of the same size in both experiments).  These speed-ups result from the reduced number of quantum circuit executions, as discussed in the subsection \ref{qdcnn}. Take Experiment A as an example.  In this experiment, $13\times13$ quantum circuits need to be executed for both the QDC layers with dilation rate $=2$ and dilation rate $=3$ while  $14\times14$ quantum circuits for standard quantum convolutional layer. This means that QDCNN\_r2 and  QDCNN\_r3 require 27 fewer quantum circuit executions per image than QCNN models. Compared with Experiment A, it generally takes much longer time to train hybrid models in Experiment B, even though the quantum filter in each of them has only eight trainable parameters coming from the random circuit. This is mainly due to the fact that PennyLane does not support vectorization for quantum circuit executions. Nevertheless, quantum dilated convolution can still help reduce the training time significantly in this case. 

Furthermore, it can also be seen from Table \ref{table1} and Table \ref{table2} that our QDCNN models generally enjoy higher recognition accuracy than the QCNN model. In particular, QDCNN\_r3  achieves the best performance with regard to both validation loss and accuracy across all tasks. In light of the QDC layer with dilation rate of 3, QDCNN\_r3 provides  up to 31.74\% lower validation loss and up to 3\%  higher validation accuracy, compared with the QCNN model. This observed model performance boost mainly stems from the contextual information in larger scales captured by the QDC layer.
\begin{table}[tbp] 
    \begin{center}
     \resizebox{0.45\textwidth}{!}{%
    \begin{tabular}{lcccc} 
     \toprule
      Dataset   & Method & Test acc & Test loss & Running time  \\ 
      \midrule
            \multirow{3}{*}{\texttt{MNIST}} &   \texttt{QCNN} & 88.00\% & 0.4389 &624.648s \\ 
             & \texttt{QDCNN\_r2} & 88.50\% & 0.3858 &\textbf{529.472}s \\
             & \texttt{QDCNN\_r3}  & \textbf{91.00}\% & \textbf{0.3466} &530.270s\\
             \midrule
             \multirow{3}{*}{\texttt{Fashion-MNIST}} &   \texttt{QCNN} & 78.50\% & 0.6319 &610.978s\\ 
             & \texttt{QDCNN\_r2} & 80.00\% & 0.6354  & 526.060s \\
             & \texttt{QDCNN\_r3}  & \textbf{81.00}\% & \textbf{0.6031}  & \textbf{525.182}s\\
             \bottomrule
               \end{tabular}
               }
    \end{center}
    \caption{Results of Experiment A on  MNIST and Fashion-MNIST datasets, reported for three hybrid-classical neural network models with non-trainable quantum filters. \texttt{QDCNN\_r2} and \texttt{QDCNN\_r3} are hybrid models with one QDC layer with dilation rate $r=2$ and $r=3$, respectively. \texttt{QCNN} represents the hybrid model with one standard quantum convolutional layer or equivalently QDC layer with dilation rate $r=1$.}
    \label{table1}
\end{table}

\begin{table}[tbp] 
    \begin{center}
     \resizebox{0.45\textwidth}{!}{%
    \begin{tabular}{lcccc} 
     \toprule
      Dataset   & Method & Test acc & Test loss & Running time  \\ 
      \midrule
            \multirow{3}{*}{\texttt{MNIST}} &   \texttt{QCNN} & 86.50\% & 0.5341 & 7.420h\\ 
             & \texttt{QDCNN\_r2} & 86.50\% & 0.5315 &6.436h \\
             & \texttt{QDCNN\_r3}  & \textbf{89.50}\% & \textbf{0.3646} & \textbf{6.053}h\\
             \midrule
             \multirow{3}{*}{\texttt{Fashion-MNIST}} &   \texttt{QCNN} & 79.50\% & 0.8923 &7.443h\\ 
             & \texttt{QDCNN\_r2} & 77.00\% & 1.1038 & \textbf{6.222}h \\
             & \texttt{QDCNN\_r3}  & \textbf{80.50}\% & \textbf{0.8837}  & 6.372h\\
             \bottomrule
               \end{tabular}
               }
    \end{center}
    \caption{Results of Experiment B on  MNIST and Fashion-MNIST datasets, reported for three hybrid-classical neural network models with trainable quantum filters. }
    \label{table2}
\end{table}
\vspace{6mm}

\section{Conclusion}\label{conclusion}
  In this work, we propose the QDCNN model, which adopts the idea of dilated convolution in deep learning to the quantum neural network.  We show through empirical evidence that the QDCNN model outperforms the recent QCNN method in terms of computation time and recognition accuracy. In particular, we find that the quantum dilated convolution with a larger dilation rate generally contribute to a better model performance. Dilated convolution has been extensively studied in the area of deep learning, but little work has been done to explore it in the context of quantum machine learning. Our work constitutes a first step in this direction. With the promising results on both MNIST and Fashion-MNIST datasets, our QDCNN approach deserves further investigation in the future. 

\bibliographystyle{ieeetr}
\bibliography{QDCNN}

\begin{thebibliography}{10}

\bibitem{lecun1998gradient}
Y.~LeCun, L.~Bottou, Y.~Bengio, and P.~Haffner, ``Gradient-based learning
  applied to document recognition,'' {\em Proceedings of the IEEE}, vol.~86,
  no.~11, pp.~2278--2324, 1998.

\bibitem{krizhevsky2012imagenet}
A.~Krizhevsky, I.~Sutskever, and G.~E. Hinton, ``Imagenet classification with
  deep convolutional neural networks,'' {\em Advances in neural information
  processing systems}, vol.~25, pp.~1097--1105, 2012.

\bibitem{simonyan2014very}
K.~Simonyan and A.~Zisserman, ``Very deep convolutional networks for
  large-scale image recognition,'' {\em arXiv preprint arXiv:1409.1556}, 2014.

\bibitem{szegedy2015going}
C.~Szegedy, W.~Liu, Y.~Jia, P.~Sermanet, S.~Reed, D.~Anguelov, D.~Erhan,
  V.~Vanhoucke, and A.~Rabinovich, ``Going deeper with convolutions,'' in {\em
  Proceedings of the IEEE conference on computer vision and pattern
  recognition}, pp.~1--9, 2015.

\bibitem{he2016deep}
K.~He, X.~Zhang, S.~Ren, and J.~Sun, ``Deep residual learning for image
  recognition,'' in {\em Proceedings of the IEEE conference on computer vision
  and pattern recognition}, pp.~770--778, 2016.

\bibitem{girshick2014rich}
R.~Girshick, J.~Donahue, T.~Darrell, and J.~Malik, ``Rich feature hierarchies
  for accurate object detection and semantic segmentation,'' in {\em
  Proceedings of the IEEE conference on computer vision and pattern
  recognition}, pp.~580--587, 2014.

\bibitem{girshick2015fast}
R.~Girshick, ``Fast r-cnn,'' in {\em Proceedings of the IEEE international
  conference on computer vision}, pp.~1440--1448, 2015.

\bibitem{redmon2016you}
J.~Redmon, S.~Divvala, R.~Girshick, and A.~Farhadi, ``You only look once:
  Unified, real-time object detection,'' in {\em Proceedings of the IEEE
  conference on computer vision and pattern recognition}, pp.~779--788, 2016.

\bibitem{he2017mask}
K.~He, G.~Gkioxari, P.~Doll{\'a}r, and R.~Girshick, ``Mask r-cnn,'' in {\em
  Proceedings of the IEEE international conference on computer vision},
  pp.~2961--2969, 2017.

\bibitem{ronneberger2015u}
O.~Ronneberger, P.~Fischer, and T.~Brox, ``U-net: Convolutional networks for
  biomedical image segmentation,'' in {\em International Conference on Medical
  image computing and computer-assisted intervention}, pp.~234--241, Springer,
  2015.

\bibitem{chen2018encoder}
L.-C. Chen, Y.~Zhu, G.~Papandreou, F.~Schroff, and H.~Adam, ``Encoder-decoder
  with atrous separable convolution for semantic image segmentation,'' in {\em
  Proceedings of the European conference on computer vision (ECCV)},
  pp.~801--818, 2018.

\bibitem{borovykh2017conditional}
A.~Borovykh, S.~Bohte, and C.~W. Oosterlee, ``Conditional time series
  forecasting with convolutional neural networks,'' {\em arXiv preprint
  arXiv:1703.04691}, 2017.

\bibitem{chen2020probabilistic}
Y.~Chen, Y.~Kang, Y.~Chen, and Z.~Wang, ``Probabilistic forecasting with
  temporal convolutional neural network,'' {\em Neurocomputing}, vol.~399,
  pp.~491--501, 2020.

\bibitem{45774}
A.~van~den Oord, S.~Dieleman, H.~Zen, K.~Simonyan, O.~Vinyals, A.~Graves,
  N.~Kalchbrenner, A.~Senior, and K.~Kavukcuoglu, ``Wavenet: A generative model
  for raw audio,'' in {\em Arxiv}, 2016.

\bibitem{yuan2019simple}
F.~Yuan, A.~Karatzoglou, I.~Arapakis, J.~M. Jose, and X.~He, ``A simple
  convolutional generative network for next item recommendation,'' in {\em
  Proceedings of the Twelfth ACM International Conference on Web Search and
  Data Mining}, pp.~582--590, 2019.

\bibitem{varatharajan2018big}
R.~Varatharajan, G.~Manogaran, and M.~Priyan, ``A big data classification
  approach using lda with an enhanced svm method for ecg signals in cloud
  computing,'' {\em Multimedia Tools and Applications}, vol.~77, no.~8,
  pp.~10195--10215, 2018.

\bibitem{kerenidis2018q}
I.~Kerenidis, J.~Landman, A.~Luongo, and A.~Prakash, ``q-means: A quantum
  algorithm for unsupervised machine learning,'' {\em arXiv preprint
  arXiv:1812.03584}, 2018.

\bibitem{otterbach2017unsupervised}
J.~Otterbach, R.~Manenti, N.~Alidoust, A.~Bestwick, M.~Block, B.~Bloom,
  S.~Caldwell, N.~Didier, E.~S. Fried, S.~Hong, {\em et~al.}, ``Unsupervised
  machine learning on a hybrid quantum computer,'' {\em arXiv preprint
  arXiv:1712.05771}, 2017.

\bibitem{article}
S.~Lloyd, M.~Mohseni, and P.~Rebentrost, ``Quantum principal component
  analysis,'' {\em Nature Physics}, vol.~10, 07 2013.

\bibitem{liu2019hybrid}
J.~Liu, K.~H. Lim, K.~L. Wood, W.~Huang, C.~Guo, and H.-L. Huang, ``Hybrid
  quantum-classical convolutional neural networks,'' {\em arXiv preprint
  arXiv:1911.02998}, 2019.

\bibitem{cong2019quantum}
I.~Cong, S.~Choi, and M.~D. Lukin, ``Quantum convolutional neural networks,''
  {\em Nature Physics}, vol.~15, no.~12, pp.~1273--1278, 2019.

\bibitem{henderson2020quanvolutional}
M.~Henderson, S.~Shakya, S.~Pradhan, and T.~Cook, ``Quanvolutional neural
  networks: powering image recognition with quantum circuits,'' {\em Quantum
  Machine Intelligence}, vol.~2, no.~1, pp.~1--9, 2020.

\bibitem{oh2020tutorial}
S.~Oh, J.~Choi, and J.~Kim, ``A tutorial on quantum convolutional neural
  networks (qcnn),'' in {\em 2020 International Conference on Information and
  Communication Technology Convergence (ICTC)}, pp.~236--239, IEEE, 2020.

\bibitem{chen2020quantum}
S.~Y.-C. Chen, T.-C. Wei, C.~Zhang, H.~Yu, and S.~Yoo, ``Quantum convolutional
  neural networks for high energy physics data analysis,'' {\em arXiv preprint
  arXiv:2012.12177}, 2020.

\bibitem{houssein2021hybrid}
E.~H. Houssein, Z.~Abohashima, M.~Elhoseny, and W.~M. Mohamed, ``Hybrid quantum
  convolutional neural networks model for covid-19 prediction using chest x-ray
  images,'' {\em arXiv preprint arXiv:2102.06535}, 2021.

\bibitem{alam2021iccad}
M.~Alam, S.~Kundu, R.~O. Topaloglu, and S.~Ghosh, ``Iccad special session
  paper: Quantum-classical hybrid machine learning for image classification,''
  {\em arXiv preprint arXiv:2109.02862}, 2021.

\bibitem{yang2021decentralizing}
C.-H.~H. Yang, J.~Qi, S.~Y.-C. Chen, P.-Y. Chen, S.~M. Siniscalchi, X.~Ma, and
  C.-H. Lee, ``Decentralizing feature extraction with quantum convolutional
  neural network for automatic speech recognition,'' in {\em ICASSP 2021-2021
  IEEE International Conference on Acoustics, Speech and Signal Processing
  (ICASSP)}, pp.~6523--6527, IEEE, 2021.

\bibitem{mitarai2018quantum}
K.~Mitarai, M.~Negoro, M.~Kitagawa, and K.~Fujii, ``Quantum circuit learning,''
  {\em Physical Review A}, vol.~98, no.~3, p.~032309, 2018.

\bibitem{schuld2019evaluating}
M.~Schuld, V.~Bergholm, C.~Gogolin, J.~Izaac, and N.~Killoran, ``Evaluating
  analytic gradients on quantum hardware,'' {\em Physical Review A}, vol.~99,
  no.~3, p.~032331, 2019.

\bibitem{Linnainmaa:1976}
S.~Linnainmaa, ``Taylor expansion of the accumulated rounding error,'' {\em BIT
  Numerical Mathematics}, vol.~16, no.~2, pp.~146--160, 1976.

\bibitem{Rumelhart:1986we}
D.~E. Rumelhart, G.~E. Hinton, and R.~J. Williams, ``{Learning Representations
  by Back-propagating Errors},'' {\em Nature}, vol.~323, no.~6088,
  pp.~533--536, 1986.

\bibitem{jones2020efficient}
T.~Jones and J.~Gacon, ``Efficient calculation of gradients in classical
  simulations of variational quantum algorithms,'' {\em arXiv preprint
  arXiv:2009.02823}, 2020.

\bibitem{Chellapilla_highperformance}
K.~Chellapilla, S.~Puri, and P.~Simard, ``High performance convolutional neural
  networks for document processing,'' in {\em Tenth International Workshop on
  Frontiers in Handwriting Recognition}, 2006.

\bibitem{vectorization}
J.~Ren and L.~Xu, ``On vectorization of deep convolutional neural networks for
  vision tasks,'' 01 2015.

\bibitem{pramanik2021quantum}
S.~Pramanik, M.~G. Chandra, C.~Sridhar, A.~Kulkarni, P.~Sahoo, V.~C. DV,
  H.~Sharma, A.~Paliwal, V.~Navelkar, S.~Poojary, {\em et~al.}, ``A
  quantum-classical hybrid method for image classification and segmentation,''
  {\em arXiv preprint arXiv:2109.14431}, 2021.

\bibitem{hur2021quantum}
T.~Hur, L.~Kim, and D.~K. Park, ``Quantum convolutional neural network for
  classical data classification,'' {\em arXiv preprint arXiv:2108.00661}, 2021.

\bibitem{schuld2019quantum}
M.~Schuld and N.~Killoran, ``Quantum machine learning in feature hilbert
  spaces,'' {\em Physical review letters}, vol.~122, no.~4, p.~040504, 2019.

\bibitem{mattern2021variational}
D.~Mattern, D.~Martyniuk, H.~Willems, F.~Bergmann, and A.~Paschke,
  ``Variational quanvolutional neural networks with enhanced image encoding,''
  {\em arXiv preprint arXiv:2106.07327}, 2021.

\bibitem{schuld2018supervised}
M.~Schuld and F.~Petruccione, {\em Supervised learning with quantum computers},
  vol.~17.
\newblock Springer, 2018.

\bibitem{larose2020robust}
R.~LaRose and B.~Coyle, ``Robust data encodings for quantum classifiers,'' {\em
  Physical Review A}, vol.~102, no.~3, p.~032420, 2020.

\bibitem{Holschneider1989}
M.~Holschneider, R.~Kronland-Martinet, J.~Morlet, and P.~Tchamitchian, ``A
  real-time algorithm for signal analysis with the help of the wavelet
  transform,'' {\em Wavelets, Time-Frequency Methods and Phase Space}, vol.~-1,
  p.~286, 01 1989.

\bibitem{Yu2016MultiScaleCA}
F.~Yu and V.~Koltun, ``Multi-scale context aggregation by dilated
  convolutions,'' {\em CoRR}, vol.~abs/1511.07122, 2016.

\bibitem{Chen2015SemanticIS}
L.-C. Chen, G.~Papandreou, I.~Kokkinos, K.~P. Murphy, and A.~L. Yuille,
  ``Semantic image segmentation with deep convolutional nets and fully
  connected crfs,'' {\em CoRR}, vol.~abs/1412.7062, 2015.

\bibitem{chen2017deeplab}
L.-C. Chen, G.~Papandreou, I.~Kokkinos, K.~Murphy, and A.~L. Yuille, ``Deeplab:
  Semantic image segmentation with deep convolutional nets, atrous convolution,
  and fully connected crfs,'' {\em IEEE transactions on pattern analysis and
  machine intelligence}, vol.~40, no.~4, pp.~834--848, 2017.

\bibitem{chen2017rethinking}
L.-C. Chen, G.~Papandreou, F.~Schroff, and H.~Adam, ``Rethinking atrous
  convolution for semantic image segmentation,'' {\em arXiv preprint
  arXiv:1706.05587}, 2017.

\bibitem{hamaguchi2018effective}
R.~Hamaguchi, A.~Fujita, K.~Nemoto, T.~Imaizumi, and S.~Hikosaka, ``Effective
  use of dilated convolutions for segmenting small object instances in remote
  sensing imagery,'' in {\em 2018 IEEE winter conference on applications of
  computer vision (WACV)}, pp.~1442--1450, IEEE, 2018.

\bibitem{kudo2017dilated}
Y.~Kudo and Y.~Aoki, ``Dilated convolutions for image classification and object
  localization,'' in {\em 2017 Fifteenth IAPR International Conference on
  Machine Vision Applications (MVA)}, pp.~452--455, IEEE, 2017.

\bibitem{chen2019environmental}
Y.~Chen, Q.~Guo, X.~Liang, J.~Wang, and Y.~Qian, ``Environmental sound
  classification with dilated convolutions,'' {\em Applied Acoustics},
  vol.~148, pp.~123--132, 2019.

\bibitem{dataencoding}
M.~Weigold, J.~Barzen, F.~Leymann, and M.~Salm, ``Data encoding patterns for
  quantum computing,'' in {\em HILLSIDE Proc. of Conf. on Pattern Lang. of
  Prog. 22}, 2019.

\bibitem{deng2012mnist}
L.~Deng, ``The mnist database of handwritten digit images for machine learning
  research [best of the web],'' {\em IEEE Signal Processing Magazine}, vol.~29,
  no.~6, pp.~141--142, 2012.

\bibitem{xiao2017fashion}
H.~Xiao, K.~Rasul, and R.~Vollgraf, ``Fashion-mnist: a novel image dataset for
  benchmarking machine learning algorithms,'' {\em arXiv preprint
  arXiv:1708.07747}, 2017.

\bibitem{Pennylane-QCNN}
A.~Mari,
  ``\href{https://pennylane.ai/qml/demos/tutorial_quanvolution.html}{Quanvolutional
  Neural Networks—PennyLane},'' 2021.

\bibitem{bergholm2018pennylane}
V.~Bergholm, J.~Izaac, M.~Schuld, C.~Gogolin, M.~S. Alam, S.~Ahmed, J.~M.
  Arrazola, C.~Blank, A.~Delgado, S.~Jahangiri, {\em et~al.}, ``Pennylane:
  Automatic differentiation of hybrid quantum-classical computations,'' {\em
  arXiv preprint arXiv:1811.04968}, 2018.

\bibitem{suzuki2011qulacs}
Y.~Suzuki, Y.~Kawase, Y.~Masumura, Y.~Hiraga, M.~Nakadai, J.~Chen,
  K.~Nakanishi, K.~Mitarai, R.~Imai, S.~Tamiya, {\em et~al.}, ``Qulacs: A fast
  and versatile quantum circuit simulator for research purpose. arxiv 2020,''
  {\em arXiv preprint arXiv:2011.13524}.

\bibitem{paszke2019pytorch}
A.~Paszke, S.~Gross, F.~Massa, A.~Lerer, J.~Bradbury, G.~Chanan, T.~Killeen,
  Z.~Lin, N.~Gimelshein, L.~Antiga, {\em et~al.}, ``Pytorch: An imperative
  style, high-performance deep learning library,'' {\em Advances in neural
  information processing systems}, vol.~32, pp.~8026--8037, 2019.

\bibitem{abadi2016tensorflow}
M.~Abadi, P.~Barham, J.~Chen, Z.~Chen, A.~Davis, J.~Dean, M.~Devin,
  S.~Ghemawat, G.~Irving, M.~Isard, {\em et~al.}, ``Tensorflow: A system for
  large-scale machine learning,'' in {\em 12th $\{$USENIX$\}$ symposium on
  operating systems design and implementation ($\{$OSDI$\}$ 16)}, pp.~265--283,
  2016.

\bibitem{Qulacs}
Qunasys, ``\href{https://github.com/qulacs/qulacs}{Qulacs},'' 2019.

\bibitem{Pennylane-Qulacs}
S.~Oud, ``\href{https://github.com/soudy/pennylane-qulacs}{PennyLane Qulacs
  plugin},'' 2019.

\end{thebibliography}

\end{document}